\documentclass{pasj00}

\usepackage{graphicx}
\usepackage{color}

\begin{document}
\SetRunningHead{Saitoh et al.}{Flaring up of G2 during the Close Encounter with Sgr A*}
\Received{2013 May 27}
\Accepted{2013 Aug 6}

\title{Flaring up of the Compact Cloud G2 during the Close Encounter with Sgr A*}

\author{Takayuki \textsc{R.Saitoh}$^{1}$
Junichiro \textsc{Makino}$^{1,2}$,
Yoshiharu \textsc{Asaki}$^{3}$,
Junichi \textsc{Baba}$^{1}$,
Shinya \textsc{Komugi}$^{4}$,
Makoto \textsc{Miyoshi}$^{4}$, 
Tohru \textsc{Nagao}$^{5}$,
Masaaki \textsc{Takahashi}$^{6}$, 
Takaaki \textsc{Takeda}$^{4,7}$, 
Masato \textsc{Tsuboi}$^{3}$, 
\& Ken-ichi \textsc{Wakamatsu}$^{8}$
}

\affil{$^1$ Earth-Life Science Institute, Tokyo Institute of
Technology, 2--12--1, Ookayama, Meguro, Tokyo, 152-8551, Japan}
\affil{$^2$ RIKEN Advanced Institute for Computational Science, 7--1--26,
Minatojima-minami-machi, Chuo-ku, Kobe, Hyogo 650-0047, Japan}
\affil{$^3$ Japan Aerospace Exploration Agency, 3--1--1, Yoshinodai,
Chuo-ku, Sagamihara, Kanagawa 252-5210, Japan}
\affil{$^4$ National Astronomical Observatory of Japan, 2--21--1,
Oosawa, Mitaka, Tokyo 182-8588, Japan}
\affil{$^5$ Kyoto University,Kitashirakawa-Oiwake-cho, Sakyo-ku, Kyoto
606-8502, Japan}
\affil{$^6$ Aichi University of Education,1 Hirosawa, Igaya-cho, Kariya,
Aichi 448-8542, Japan}
\affil{$^7$ VASA Entertainment Co., Ltd., 4-14-17, Senninchou, Hachioji, Tokyo,
193-0835, Japan}
\affil{$^8$ Gifu University,1--1 Yanagido, Gifu City 501-1193, Japan}
\email{saitoh@geo.titech.ac.jp}


\maketitle 

\begin{abstract}
A compact gas cloud G2 is predicted to reach the pericenter of its orbit around
the super massive black hole (SMBH) of our galaxy, Sagittarius A* (Sgr A*).
This event will give us a rare opportunity to observe the interaction between
SMBH and gas around it.  We report the result of the fully three-dimensional
simulation of the evolution of G2 during the first pericenter passage.  The
strong tidal force by the SMBH stretches the cloud along its orbit, and
compresses it strongly in the vertical direction, resulting in the heating up
and flaring up of the cloud.  The bolometric luminosity will reach the maximum
of $\sim100~L_{\odot}$. This flare should be easily observed in the near
infrared.
\end{abstract}

\section{Introduction} \label{sec:intro}

At present, the activity of Sgr A* seems to be in the low-state, with the X-ray
luminosity of $10^{33}~{\rm erg~s^{-1}}$ \citep{Baganoff+2003}. There are
evidences of past activities \citep{Sunyaev+1993, Koyama+1996, Dobler+2010,
Su+2010}, where the luminosity had reached as high as $\sim 10^{40}~{\rm
erg~s^{-1}}$.  Recently, rapid flaring from Sgr A* was observed in various wave
lengths \citep{Tsuboi+1999, Baganoff+2001, Genzel+2003, Miyazaki+2004}.  Thus,
it is quite important to understand how these large variations in the luminosity
took place.  One possibility is the intermittent supply of gas in the form of
high-density clouds.

The compact cloud G2 \citep{Gillessen+2012} might be offering us the first
opportunity to study such an interaction of a gas cloud and SMBH.  While the
formation mechanism of the cloud is under debate \citep{Burkert+2012,
Gillessen+2012, Miralda-Escude2012, Murray-ClayLoeb2012,
MeyerMeyer-Hofmeister2012, ScovilleBurkert2013}, we know that the orbit of G2,
determined from observations since 2004, is highly eccentric, and G2 will reach
the pericenter of its orbit in the mid 2013 ($2013.51\pm0.04$), with the
pericenter distance of only $270~{\rm au}$ \citep{Gillessen+2012}.
\citet{Gillessen+2013} reported the updates of pericentric distance and
pericenter passage epoch as $190~{\rm au}$ and Sep 2013 ($2013.69\pm0.04$),
respectively.  \citet{Phifer+2013} reported different values of $130~{\rm au}$
and Mar 2014 ($2014.21\pm0.14$).  In \citet{Gillessen+2013b}, these are now
$210~{\rm au}$ and Mar 2014 ($2014.25\pm0.06$).

\citet{Schartmann+2012} studied the evolution of G2 using high-resolution
adoptive mesh refinement calculation in two dimensions.  In their calculation,
the cloud loses the kinetic energy during the pericenter passage due to the ram
pressure from the hot atmosphere around the SMBH, and gas accretion to SMBH
starts in early 2013, and continues for several decades with a nearly constant
accretion rate.

However, it is not clear if the two-dimensional calculation is appropriate or
not. The cloud should experience strong compression in the direction
perpendicular to the orbital plane, due to the tidal force from the SMBH,
resulting in a very thin structure.  Because of this structure change, the ram
pressure might become ineffective unlike two-dimensional simulations.  In addition,
this compression energy is emitted via radiation immediately. It is therefore
expected that the luminosity of the cloud will increase during the pericenter
passage.  Since the tidal force in the vertical direction is proportional to the
distance from the orbital plane, the cloud will contract uniformly.  There is no
shock during this contraction, as long as the cloud maintains a finite
thickness.  \citet{Anninos+2012} carried out the three-dimensional mesh
simulations, but they neglected the effects of the radiative cooling and
therefore did not notice this brightening.

In order to study these effects, we performed fully three-dimensional
simulations, in which the compressional heating and radiative cooling of the
cloud are consistently taken into account.

\section{Method}\label{sec:Method}

We solved the evolution of a system consisting of Sgr A*, hot-ambient gas, and
the cloud by $N$-body/Smoothed Particle Hydrodynamics (SPH) simulations.  Here,
we adopted the compact cloud scenario \citep{Burkert+2012}. 

We modelled Sgr A* as a sink particle \citep{BateBurkert1997} with the mass of
$4.31\times10^6~M_{\odot}$ \citep{Gillessen+2009}.  This sink particle can
absorb nearby gas particles. The sink radius is $30~{\rm au}$, which is 350
times larger than the real horizon scale of SMBH, $0.085~{\rm au}$, and is 10
times smaller than the pericenter distance.  When a gas particle is absorbed by
the sink particle, the gas particle is removed and its mass is added to that of
the sink particle. We did not consider the emission from the absorbed gas since
observations suggest that accretion flow around Sgr A* is expressed by
radiatively inefficient accretion flow (RIAF) \citep{Ichimaru1977,
NarayanYi1994}.

A diffuse and hot X-ray emitting gas around Sgr A* \citep{Yuan+2003, Xu+2006}
was modelled by Yuan's RIAF model \citep{Yuan+2003}, following previous studies
of G2 \citep{Gillessen+2012, Burkert+2012, Schartmann+2012}.  The density and
temperature profiles are given by
\begin{eqnarray}
\rho_{\rm hot}(r)&=&1.7\times10^{-21}f_{\rm hot}
\left(\frac{1.0\times10^{16}~{\rm cm}}{r}\right)~{\rm g~cm^{-3}}\label{eq:Density},\\
T_{\rm hot}(r)&=&2.1\times10^8\left(\frac{1.0\times10^{16}~{\rm cm}}{r}\right)~{\rm K},
\label{eq:Temperature}
\end{eqnarray}
where $r$ is the distance from Sgr A* and $f_{\rm hot}$ is the scaling parameter
of gas density. We changed $f_{\rm hot}$ from $1.0$ (Run 1) through $0.1$ (Run 2)
to $0$ (Run 3) in order to investigate the effect of the hot gas to the
evolution of G2.  The rotation and inhomogeneity of the hot gas at the initial
state were neglected, whereas the dynamical evolution was allowed.  Although
this profile is convectively unstable \citep{Schartmann+2012}, we did not try to
prevent the growth of convection. In our model, we allowed the radiative cooling
of the hot gas, resulting in the accretion rate consistent to the value
suggested by the observation. Because of this accretion flow, the growth of the
convection was effectively suppressed.

The cloud, G2, was modelled as a spherical gas cloud of three earth mass and a
uniform density distribution.  The initial radius of the cloud is $125~{\rm au}$
and the initial temperature of the gas is $10^4~{\rm K}$.  The orbit of the
cloud is that of \citet{Gillessen+2012} where the pericenter distance is
$270~{\rm au}$ and the pericenter passage epoch is $2013.5$.  We adopted A.D.
1995 as the starting epoch of the simulations and solved the evolution of the
cloud for $38$ years.  This cloud was in the hydrostatic equilibrium with the
ambient gas in A.D. 1995, $r_{1995} \simeq 5100~{\rm au}$: the pressure ratio of
the cloud to the hot ambient gas was $1.1/f_{\rm hot}$.  If the cloud is formed
at the apocenter, it should has too elongated structure which is inconsistent to
the observation \citep{Burkert+2012}.

Particle number, mass and spatial resolutions of the cloud, hot ambient, and
SMBH are summarized in Table \ref{tab:particles}.

\begin{table*}[htb]
\begin{center}
\caption{Number of particles and mass and spatial resolutions}\label{tab:particles}
\begin{tabular}{rccc}
\hline
\hline
Component&Number&Particle mass&Softening length\\
\hline
Cloud (Run 1)&$1\times10^6$&$3\times10^{-6}~M_{\oplus}$&$0.43~{\rm au}$\\
Cloud (Run 2)&$3\times10^5$&$1\times10^{-5}~M_{\oplus}$&$0.65~{\rm au}$\\
Cloud (Run 3)&$10^7$ & $3\times10^{-7}~M_{\oplus}$&$0.20~{\rm au}$\\
\hline
Hot gas (Run 1)&$1\times10^7$&$2.8\times10^{-5}~M_{\oplus}$&$0.92~{\rm au}$\\
Hot gas (Run 2)&$3\times10^6$&$9.4\times10^{-5}~M_{\oplus}$&$0.63~{\rm au}$ \\
Hot gas (Run 3)&N/A&N/A&N/A\\
\hline
SMBH (Run 1,2,3)&$1$&$4.31\times10^{6}~M_{\odot}$&$10~{\rm au}$ \\
\hline
\end{tabular}\\
\end{center}
\end{table*}

We used {\tt ASURA}, a parallel $N$-body/SPH simulation code, for these
simulations \citep{Saitoh+2008, Saitoh+2009}.  Gravity was solved by the tree
with GRAPE method \citep{Makino1991TreeWithGRAPE}.  A symmetrized potential was
used in order to accelerate the gravity calculation with tree with the
individual softening length \citep{SaitohMakino2012}.  In this study, we adopted
the density independent SPH \citep{SaitohMakino2013} in which the pressure, or
the energy density, is evaluated first, and other quantities are evaluated using
the pressure. This formulation can successfully handle hydrodynamical
instabilities. This ability would be important since according to
\citet{Burkert+2012}, the hydrodynamical instabilities might play important
roles in the cloud evolution, in particular at the pericenter.  We used the
second order symplectic integrator, the leap-frog method, and the individual
time-step method \citep{McMillan1986, Makino1991IndividualTimeStep}.  The FAST
method \citep{SaitohMakino2010} and the time-step limiter
\citep{SaitohMakino2009} were also used.

We adopted the Monaghan type artificial viscosity term \citep{Monaghan1997} to
handle the shock.  To avoid the penetration of particles in the vertical
direction at the pericenter passage, we adopted a rather large value of the
viscosity parameter, $\alpha=6$.  The radiative cooling and the photo-electric
heating due to the far-ultraviolet (FUV) field were taken into account in the
form of a cooling/heating function \citep{Wada+2009, Wolfire+1995} and an
optically thin approximation was used.  With this function, the FUV heating is
modelled through the heating rate $G_0$ (see appendix B in \cite{Wada+2009}).
The covered range of $G_0$ is $0-10^4$, which corresponds to $0-6,000$ times as
that at the solar neighborhood.  On the other hand, if we assume that the FUV
heating is proportional to the local stellar density, it is $10^7$ times as that
of the solar neighborhood.  The stellar density at the galactic center is $\sim
10^6~{\rm M_{\odot}~pc^{-3}}$ \citep{Genzel+1996, Haller+1996} and that at the
solar neighborhood is  $\sim 0.05 ~{\rm M_{\odot}~pc^{-3}}$ \citep{Creze+1998,
HolmbergFlynn2000}.  Therefore, we could not give FUV heating sufficient
strength.  We will discuss what is expected when we used $G_0 \sim 10^7$ in \S
\ref{sec:Heating}.  We assumed $2~Z_{\odot}$ for gas, since the metallicity
fraction at the galactic central region is $1.5-3~Z_{\odot}$
\citep{Genzel+2010}.

Since the radiative cooling is very strong, the temperature of the gas cloud is
always less than $10^4~{\rm K}$. This is the reason why previous studies
excluded the radiative cooling and often assumed adiabatic or isothermal EOS
\citep{Burkert+2012, Schartmann+2012, Anninos+2012}. However, as we show in this
paper, to include the effect of the radiative cooling is quite important to
predict the evolution of luminosity.

\section{Results} \label{sec:Results}

\begin{figure*}[htb]
\begin{center}
\includegraphics[width=0.9\textwidth]{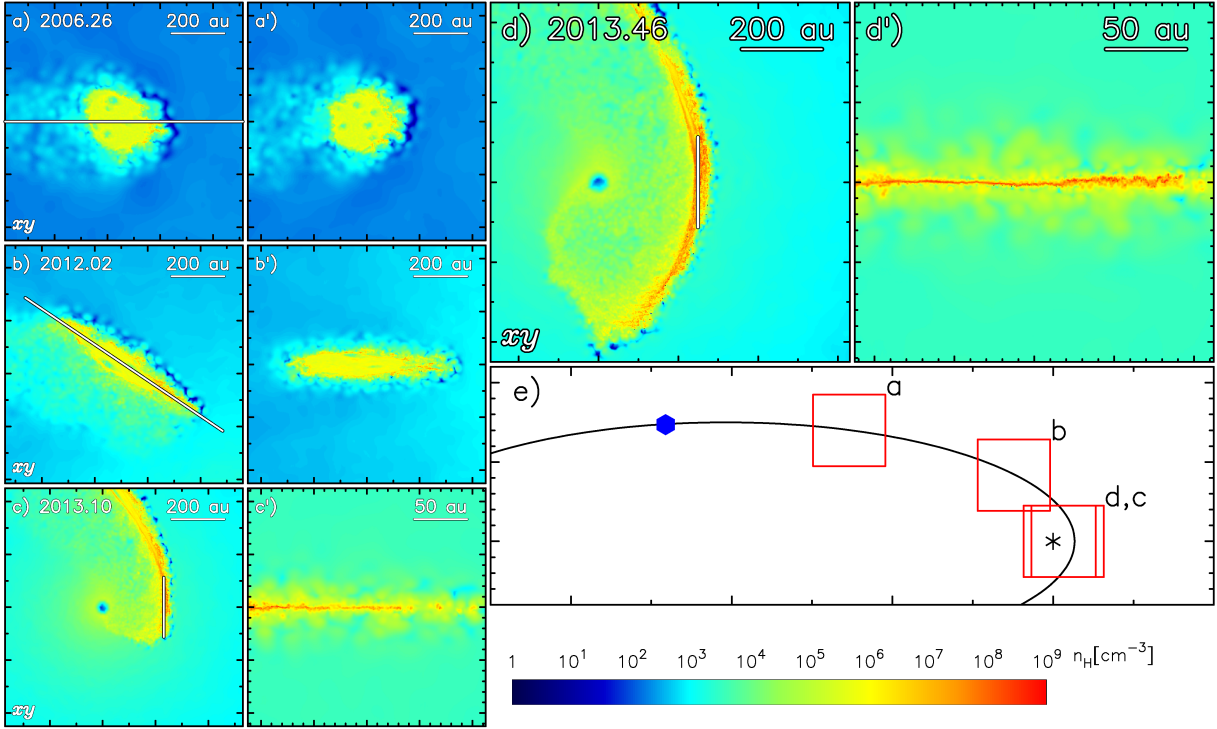}
\caption{
Time evolution of the structure of G2 for the period of A.D. 2006.26 to A.D.
2013.46, tagged with a to d and a' to d'.  The density distributions in the $xy$
planes (a to d) and the selected planes (a' to d') are shown.  Here, the $xy$
plane is corresponding to the orbital plane of G2.  The thin solid lines in the
panels for $xy$ planes show the location of the plane for which the density
distributions are shown in the panels in the corresponding panels for selected
planes.  The top-right two panels shows the distribution at A.D. 2013.46. The
left two columns show the view in $xy$ and selected planes, respectively. The
panel e depicts part of the orbit of G2 (solid curve) and the positions of the
$xy$ planes in the four epochs (red squares).  The blue hexagon indicates the
position of the cloud at A.D. 1995.5. The plotted region is $-7000~{\rm
au}<x<2000~{\rm au}$ and $-800~{\rm au}<y<2200~{\rm au}$.}\label{fig:Snapshots} 
\end{center} 
\end{figure*}

Figure \ref{fig:Snapshots} shows the evolution of the three-dimensional
structure of the cloud from A.D. 2006.26 to A.D. 2013.46.  For this model, we
used $f_{\rm hot}=1$ which is similar to that used in the previous
two-dimensional calculation \citep{Burkert+2012, Schartmann+2012}.  At A.D.
2006, the simulated G2 was nearly spherical, since the effect of the tidal force
is inefficient. The destruction effect due to hydrodynamical instabilities is
also inefficient since the time scale of instabilities is sufficiently long,
$\sim10~{\rm yr}$ \citep{Gillessen+2012}.  By A.D. 2012.02, it is stretched in
the orbital plane and compressed in the vertical direction.  This stretch has
already been observed \citep{Gillessen+2012}.  When the cloud passes the
pericenter, its thickness reaches the minimum. Due to this strong vertical
compression, the gas density increases by more than two orders of magnitude.

In the last panels (panels d and d'), we can see the ``bridge'' between the
central SMBH and the head of the cloud. This bridge indicates that there is a
flow of gas from the head of the cloud to the SMBH. However, the amount of the
gas in this bridge, and the resulting accretion rate to SMBH, are small.

\begin{figure}
\begin{center}
\includegraphics[width=0.4\textwidth]{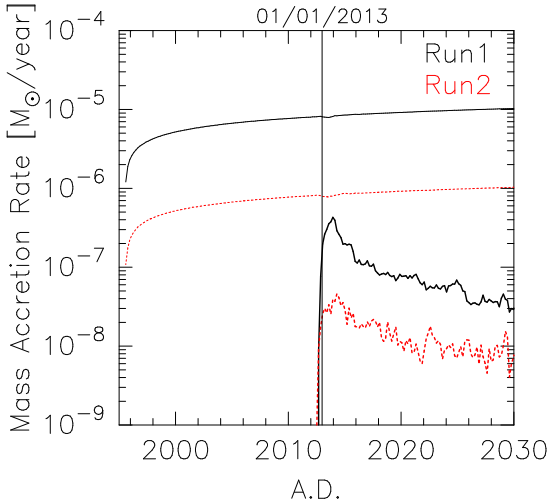}
\caption{
Time evolution of the accretion rate of the gas to the SMBH.  Black and red
curves are the results of simulations with the standard high-density hot
atmosphere (Run1) and with low-density atmosphere (Run2), respectively.  Thick
curves show accretion rates of the cloud gas, and thin curves show those of the
halo gas.  The vertical line denotes the epoch of 01/01/2013.  When we excluded
the hot ambient, there was no accretion to the SMBH.
}\label{fig:AccretionRate}
\end{center}
\end{figure}

Figure \ref{fig:AccretionRate} shows the history of accretion rates. We
show the results from two runs out of the three runs. For the first run, we used
the cloud model the same as that used in \citet{Burkert+2012} and
\citet{Schartmann+2012}, except that we solve the dynamical evolution of the hot
ambient.  Whether or not such high-density, high-temperature atmosphere actually
exists near SMBH is an open question. In order to test the importance of the
assumption on the hot atmosphere, we performed two additional simulations. In
the second run, we reduced the gas density by a factor of 10 (red dashed curve
in the figure), and in the third run we eliminated the atmosphere altogether.

In the case of the standard run, the accretion rate of the gas from the cloud
reaches the peak value of $\sim10^{-7}~{\rm M_{\odot}~year^{-1}}$ at A.D.
2014, and then decreases exponentially. The accretion rate at A.D. 2030 is one
order of magnitude smaller than the peak value.

These behaviours are quite different from those in two-dimensional calculations
\citep{Schartmann+2012}, in which the accretion rate is nearly constant due to
the strong ram pressure. In our three-dimensional simulations, the height of the
cloud is reduced to 1/100 of the initial value at the pericenter, resulting in
the decrease of the effect of the ram pressure by a similar factor (see
\S\ref{sec:rampressure}).  The total amount of the mass accretion from the gas
cloud till A.D. 2023.5, the first ten years from the pericenter passage, is
$15\%$ of the cloud mass for Run 1. This accretion rate of Run 1 is comparable
to that obtained in a two-dimensional simulation \citep{Schartmann+2012}.  Since
the gas cloud in our three-dimensional simulation is vertically compressed, one
might expect a much lower accretion rate.  The main reason for this high
accretion rate is that in our model the ``ambient'' gas is accreted to the SMBH,
carrying the gas removed from the cloud.  In the two-dimensional simulation of
\citet{Schartmann+2012}, the ambient gas is pinned to the original position.
Note that the higher accretion rate of the hot ambient gas indicates that, even
if we assumed a relatively high density for the halo gas, it is probably
difficult to observe the change of the activity of Sgr A* due to this additional
accretion.

The overall evolution of the accretion rate depends strongly on the assumed
density of the hot atmosphere.  When we reduced the hot gas density by a factor
of 10, the accretion rate decreased by the same factor.  The total amount of the
accreted mass during the first ten years from the pericenter passage is
$\sim2\%$ of the original cloud mass.  In the run with no halo gas, no gas is
accreted to the SMBH.

\begin{figure*}
\begin{center}
\includegraphics[width=0.9\textwidth]{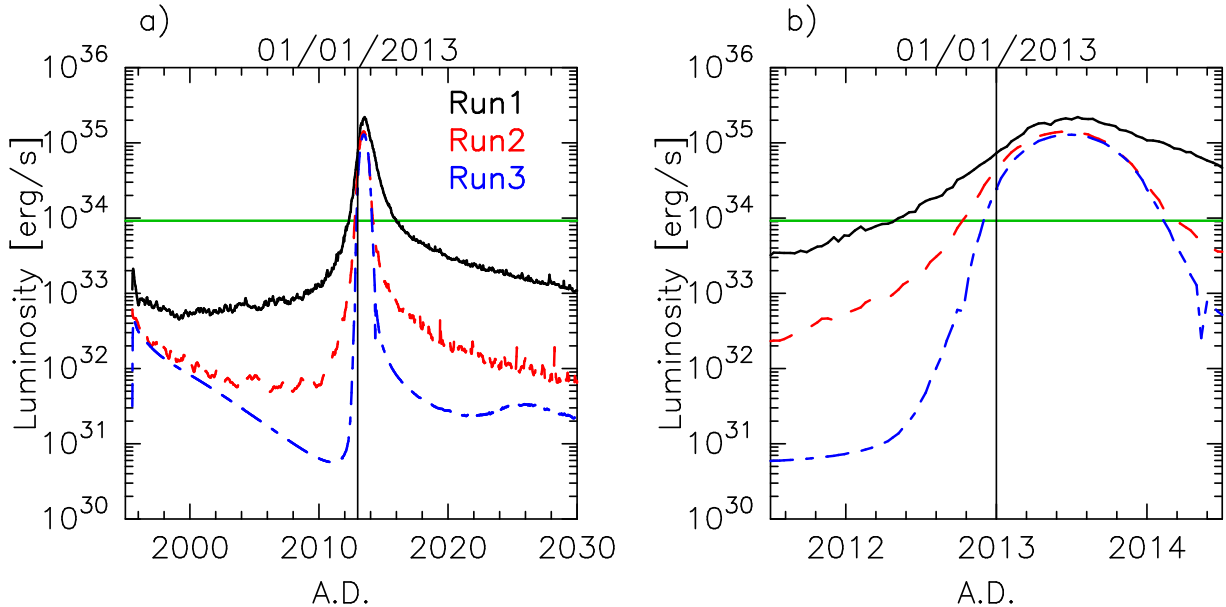}
\caption{
Bolometric luminosity as a function of time. Panel a shows the time evolution of
the luminosity from A.D. 1995 to A.D. 2030, while panel b shows it from A.D.
2011.5 to 2014.5. Solid black, red dashed and blue dot-dashed curves are the
results from simulations with the standard high-density atmosphere (Run1),
with low-density atmosphere (Run2), and no atmosphere (Run3), respectively.  
The green horizontal line indicates the expected luminosity of $9.2
\times 10^{33}~{\rm erg~s^{-1}}$ under the strong heating rate with $G_0 = 1.7
\times 10^7$. See the text in \S \ref{sec:Heating}.
}\label{fig:EnergyLoss}
\end{center}
\end{figure*}

Figure \ref{fig:EnergyLoss} shows the evolution of the bolometric luminosity of
the cloud, for three different models.  We integrated the cooling rate of the
gas particles during simulations. Note that we neglect the emission from the gas
absorbed by Sgr A*, because the accretion around Sgr A* is expected as RIAF.  In
all of three runs the luminosity peaks at the time of the pericenter passage.
Before the pericenter passage, the luminosity is almost constant for the
standard run, but goes down for other two runs.  In the case of the standard
run, there is friction with the hot gas supplies the thermal energy to the
cloud, and the luminosity is kept nearly constant. For other two runs, the hot
gas is much less dense and heating effect is much smaller.  However, the model
variation would vanish when we assume much stronger heating rate which is
adequate to reproduce the environment of the galactic center region (see \S
\ref{sec:Heating}).  The peak luminosity and its duration is practically
independent of the assumption for the hot atmosphere.  This result is quite
natural, since the peak luminosity comes mainly from the tidal compressional
heating of the cloud and has nothing to do with the interaction with the
atmosphere. On the other hand, the interaction with hot atmosphere keeps the
luminosity high for years before and after the pericenter passage.  We will give
rough estimate of the effect of ram-pressure heating in \S\ref{sec:rampressure}.

We can estimate the total amount of the energy generated by the tidal
compressional heating in the following way. For simplicity, let's assume that
gas moves freely until it reaches the equatorial plane, where it converts all
the kinematic energy of its vertical motion to thermal energy.  This is of
course not a realistic assumption, since the gas is heated due to the
compression and emits radiation. Therefore, strictly speaking, what is given
below is the upper bound.  The vertical velocity is given by $V_{\rm v}=V_{\rm
a} \tan i$, where $V_{\rm a}$ is the velocity at the ascending node and $i$ is the
inclination. If we assume that the cloud was still spherical around year A.D.
2000 where the cloud was around the one of vertice, we have 
\begin{equation}
V_{\rm v} \simeq 440\left(\frac{V_{\rm a}}{5300~{\rm km~{^{-1}}}}\right)
\left(\frac{R_{\rm c}}{125~{\rm au}}\right)\left(\frac{1500~{\rm au}}{R_{\rm b}}\right)~{\rm km~s^{-1}}
\label{eq:tangential_velocity}
\end{equation}
for a gas element $125~{\rm au}$ from the orbital plane at A.D. 2000.  Here,
$R_{\rm c}$ is the radius of the cloud when the cloud is on the minor axis of
its orbit and $R_{\rm b}$ is the distance to the vertice in the semi-minor axis.
By integrating the energy over the spherical cloud of radius
$R_{\rm c}$, we have
\begin{eqnarray}
\frac{dE_{\rm t}}{dt}&\simeq&2.2\times10^{35}\left(\frac{V_a}{5300~{\rm km~{^{-1}}}}\right)^2\left(\frac{1500~{\rm au}}{R_b}\right)^2\nonumber\\
&\quad\quad&\left(\frac{M_{\rm c}}{3M_{\oplus}}\right)\left(\frac{R_{\rm c}}{125~{\rm au}}\right)^2\left(\frac{1~{\rm year}}{\tau}\right)~{\rm erg~s^{-1}},\label{eq:tidalheating}
\end{eqnarray}
where $M_{\rm c}$ is the cloud mass.  We assume that vertical velocity is zero
at that moment. In other words, we assume that the ascending node coincides with
the pericenter.  The duration time of energy release $\tau$ is about one year,
as we can see from figure 3.

The radiative energy loss rate of the cloud is given by
\begin{equation}
\frac{dE_{\rm cooling}}{dt}\simeq1.1\times10^{39}\left(\frac{n_H}{10^5~{\rm cm^{-3}}}\right)\left(\frac{M_{\rm c}}{3M_{\oplus}}\right)~{\rm erg~s^{-1}}. 
\end{equation}
Here, we used the cooling coefficient of $\Lambda\sim10^{-22}~{\rm
erg~cm^3~s^{-1}}$ at the gas of $10^4~{\rm K}$ \citep{SutherlandDopita1993}.
Since the cooling rate is sufficiently large, it is possible to keep the
temperature of the cloud to $\sim10^4~{\rm K}$ (see also figure 4), and
therefore the emission is mostly in hydrogen recombination lines.

This simple estimate is in good agreement with the total amount of radiation in
our detailed simulations, although the actual evolution process would be much
complex.  Note that the total amount of radiation depends on the height and
vertical velocity structure of the cloud.  For example, if the compression
velocity is zero at the apocenter, the inclination would be smaller by a factor
of a few, resulting in the decrease of the total luminosity by about one order
of magnitude.

\begin{figure*}
\begin{center}
\includegraphics[width=0.9\textwidth]{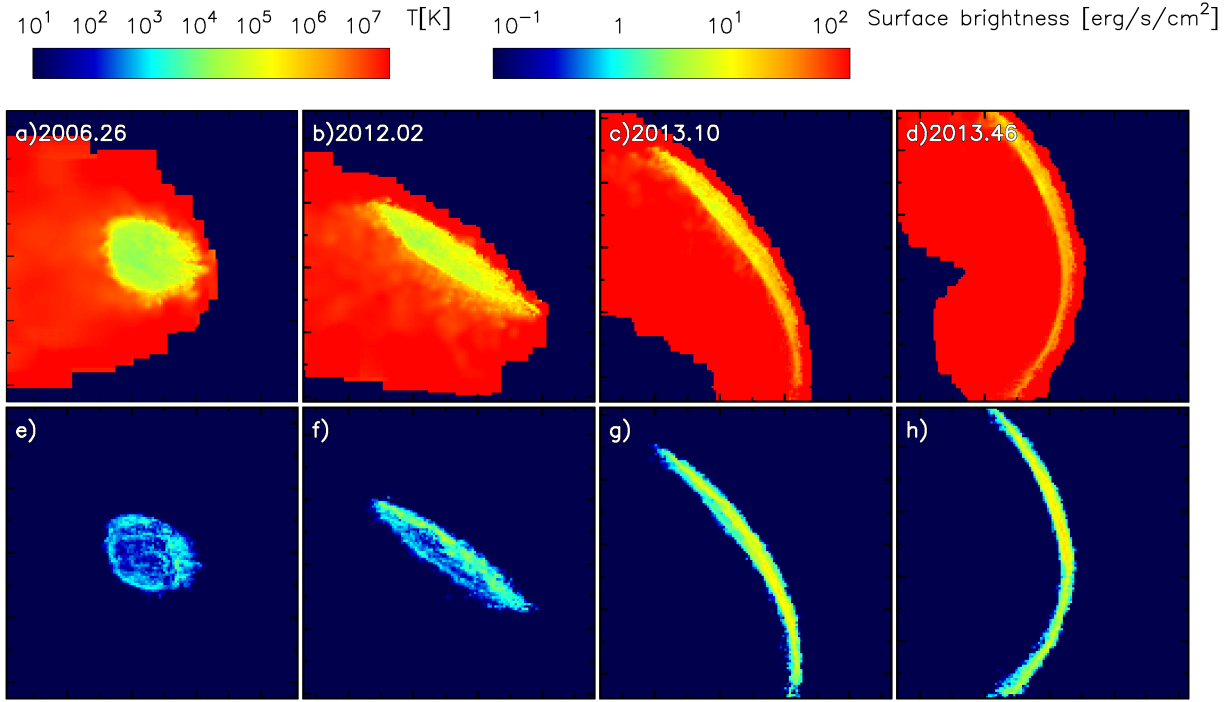}
\caption{
Time evolution of the gas temperature and integrated surface luminosity from gas
with a temperature range of $8000-15000~{\rm K}$ from A.D. 2006.26 to A.D.
2013.46, tagged with a to d and e to h. The face-on view (the $xy$ plane) is
assumed and each panel shows a $900~{\rm au}\times900~{\rm au}$
region.}\label{fig:Temperature}
\end{center}
\end{figure*}

Figure \ref{fig:Temperature} shows the distributions of temperature and
brightness, for selected moments.  Here, only the gas component which is
initially associated with the gas cloud is considered in this figure.
The main body of the gas cloud is heated up by the compression, but the
temperature remains near $10^4~{\rm K}$ due to very efficient radiative cooling
of ionized hydrogen through recombination lines. The most luminous region is
several hundred aus in size. They are most likely to be observable from the
outside in the near infrared, which does not suffer from the dust extinction
effects significantly.

\section{Discussion \& Summary} \label{sec:Discussion}

\subsection{Contribution of Ram Pressure in Three-dimensional Simulations}
\label{sec:rampressure}

The ram pressure of the hot ambient to the cloud is evaluated as
\begin{equation}
P(r)=\rho_{\rm hot}(r)v_{\rm c}(r)^2,
\end{equation}
where $\rho_{\rm hot}$ is the density of the hot ambient gas and $v_{\rm c}$ is the
velocity of the cloud relative to the hot ambient.  When we assume that some
part of the work of the pressure force is converted to the thermal energy of the
cloud, the heating rate is 
\begin{equation}
\frac{dE_{\rm ram}}{dt}= C \sigma P(r) v_{\rm c}(r) =C \sigma \rho_{\rm hot}(r)v_{\rm c}(r)^3,
\end{equation}
where $\sigma$ is the cross-section of the cloud, {\it i.e.,} the size
of the cloud projected to the plane perpendicular to its motion, and $C$ is the
conversion efficiency.  By using the density profile of the hot ambient gas (Eq.
\ref{eq:Density}) and the relation between $r$ and $v_c$ (see Eq. 1 in
\cite{Burkert+2012}), we have 
\begin{eqnarray}
\frac{dE_{\rm ram}}{dt}&\simeq&8.2\times10^{33}Cf_{\rm hot}
\left[\frac{\sigma}{\pi(125~{\rm au})^2}\right]
\left(\frac{6\times10^{16}~{\rm cm}}{r}\right)
\nonumber\\
&&\left[\left(\frac{6\times10^{16}~{\rm cm}}{r} \right)-0.46\right]^{3/2}
~{\rm erg~s^{-1}}.\label{eq:ram}
\end{eqnarray}
We can see that the heating rate depends both on the distance from the SMBH and
the cross section of the cloud, {\it i.e.,} $r$ and $\sigma$.  
Thus, to estimate the time variation of the heating due to the ram
pressure, we need to estimate the time variation of the cross-section.

As an example, we evaluate the ram-pressure heating rate at the pericenter.
From figure \ref{fig:Snapshots}, the cross-section of the cloud at the
pericentre is $\sigma_{\rm A.D. 2013.5}\sim1~{\rm au}\times40~{\rm au}=40~{\rm
au}^2$.  The thickness of $1~{\rm au}$ is affected by the resolution limit, and
in reality the cloud is probably even more thinner.  Hence, the estimate below
gives the upper limit.  Substituting this value and
$r=r_{\rm A.D. 2013.5}=4\times10^{15}~{\rm cm}$ into Eq. \ref{eq:ram}, we have
\begin{equation}
\left(\frac{dE_{\rm ram}}{dt}\right)_{\rm A.D. 2013.5}
\le5.6\times10^{33}Cf_{\rm hot}~{\rm erg~s^{-1}}.
\end{equation}
Even when we adopt $C=1$, this value is nearly two orders of magnitude smaller
than that by the tidal heating (see Eq. \ref{eq:tidalheating}).  The ram
pressure is not the primary source of the luminosity of the cloud.

As the other example, we evaluate Eq. \ref{eq:ram} at A.D. 2000, where
$r_{\rm A.D. 2000}=6\times10^{16}~{\rm cm}$.  We assume that the cloud shape
maintains the original spherical shape at this moment. Therefore, $\sigma_{\rm
A.D. 2000} \sim\pi(125~{\rm au})^2$. By substituting these values, we obtain 
\begin{equation}
\left(\frac{dE_{\rm ram}}{dt}\right)_{\rm A.D. 2000}
\simeq3.3\times10^{33}Cf_{\rm hot}~{\rm erg~s^{-1}}.
\end{equation}
This rate is about three times higher than the result of our simulation. If we
assume $C\sim0.3$, the ram-pressure heating explains the luminosity before
pericenter passage in our simulation fairly well.

\subsection{Cloud Luminosity before the Pericenter Passage}\label{sec:Heating}
As described in \S \ref{sec:Method} and \S \ref{sec:Results}, the adopted
heating rate due to FUV in our simulations was too low compared to the expected value.
Here, we discuss the expected bolometric luminosity before the pericenter
passage.

According to \citet{BakesTielens1994}, the heating rate due to the photoelectric
heating by the far-ultraviolet field is 
\begin{equation}
n_{\rm H} \Gamma = 10^{-24} n_{\rm H}  G_{0}\epsilon_{0}~{\rm erg~cm^{-3}~s^{-1}},
\end{equation}
where $G_0$ is the coefficient of the heating rate and $\epsilon_0$ is the
efficiency and we deal with this as a constant value $0.05$, although it depends
weakly on $G_0$, $T$, and $n_{\rm H}$ \citep{BakesTielens1994}.  The
heating rate of the cloud is, thus, 
\begin{equation}
\frac{dE_{\rm heating}}{dt} \simeq
5.4 \times 10^{28} 
\left ( \frac{M_{\rm c}}{3 M_{\oplus}} \right )
\left ( \frac{G_0}{10^2} \right )
\left ( \frac{\epsilon_0}{0.05} \right ) {\rm erg~s^{-1}}.
\end{equation}
As discussed in \S \ref{sec:Method}, we could not used $G_0 > 10^4$, and the
actual value we used is $10^2$.  This value is quite low and hence we cannot
observe the effect of the heating in figure \ref{fig:EnergyLoss}. 

The expected value of $G_0$ is $\sim 1.7\times10^7$, which gives the heating
rate of $9.2 \times 10^{33}~{\rm erg~s^{-1}}$.  The green horizontal line in
figure \ref{fig:EnergyLoss} indicates this luminosity.  In this case, the
heating rate is always larger than the heating by the ram pressure.  Thus, FUV
heating should be the primary source of the cloud luminosity before the
pericenter passage and all runs show the same and constant luminosity. The time
independent luminosity is consistent with the observations (See also \S
\ref{sec:BrG}).

\subsection{Luminosity in the Br$\gamma$ line}\label{sec:BrG}

Based on our simulation results, we discuss the Br$\gamma$ magnification during
the pericenter passage. Since the main cooling mechanism of the gas cloud is the
line cooling, we computed the fraction of the Br$\gamma$ line luminosity to the
total emission-line luminosity, $F$.  For this computation, we used the publicly
available code Cloudy ver. c10.00 \citep{Ferland+1998}.  We assumed a
compressed gas with the hydrogen density of $10^{6-8}~{\rm cm^{-3}}$, the solar
chemical abundances, and typical grains. We then computed the emission-line
spectrum when this gas is heated to be $2\times10^4~{\rm K}$. We found that
$F\sim0.1$\%.  

By multiplying $F$ to the bolometric luminosity, we have the Br$\gamma$
luminosity of several $\times10^{32}~{\rm erg~s^{-1}}$ at A.D.  2013.5.
According to \citet{Gillessen+2012} and \citet{Gillessen+2013}, the intrinsic
luminosity of the Br$\gamma$ line from G2 is $0.166\%-0.2\%$ of the solar
luminosity, $\sim7\times10^{30}~{\rm erg~s^{-1}}$ during A.D. 2004-2012.  Thus,
the Br$\gamma$ luminosity at the peak will reach to nearly 100 times that of
observed values before the pericenter passage.  

Applying the value of $F$ to the expected luminosity before the pericenter
passage we obtained in \S \ref{sec:Heating}, we obtain the constant Br$\gamma$
flux of $\sim 10^{-3} L_{\odot}$.  This value 
is consistent with the observational results that
the Br$\gamma$ flux is $\sim 2\times 10^{-3} L_{\odot}$ and almost constant
during this nine years since A.D. 2004 \citep{Gillessen+2012, Gillessen+2013,
Gillessen+2013b}.  

\subsection{Peak Bolometric Luminosities with Different Orbits}\label{sec:Others}

So far, four studies have reported the orbital information of G2 and there are
some variations. Here, we evaluate the peak luminosities of the cloud due to the
tidal compression in these orbits.

Table \ref{tab:Orbits} summarizes the orbital information of four studies
\citep{Gillessen+2012, Gillessen+2013, Gillessen+2013b, Phifer+2013} and
expected peak luminosities evaluated by Eq. \ref{eq:tidalheating}.
Note that we assumed the same duration time for all orbits.  From this table, we
can see that the variations of the expected peak luminosities are at most a
factor of three. The orbit reported by \citet{Phifer+2013} gives the highest
luminosity, reflecting the highest eccentricity and the closest pericentric
distance.

We also note that the time for which the cloud will stay around the pericenter
decreases when the $V_{\rm a}$ increases. As a result, the rise and decay of the
light curve become steeper and the duration time decreases.  This should change
the peak luminosity of the light curve, but we do not take this point into
account in this table.  Simulations with different orbits are necessary to have
concrete expectations and we will show them in the near future.

\begin{table*}[htb]
\begin{center}
\caption{Orbital Information and Expected Peak Luminosities}\label{tab:Orbits}
\begin{tabular}{rcccc}
\hline
\hline
&\citet{Gillessen+2012}&\citet{Gillessen+2013}&\citet{Gillessen+2013b} &
\citet{Phifer+2013}\\
\hline
${r_{\rm peri}}^{*}$ & $270~{\rm au}$ & $190~{\rm au}$ & $210~{\rm au}$ & $130~{\rm au}$ \\
${e}^{\dagger}$ & $0.938$ & $0.966$ & $0.976$ & $0.981$ \\
${R_{\rm a}}^{\ddagger}$ & $4300~{\rm au}$ & $5240~{\rm au}$ & $8750~{\rm au}$ & $6840~{\rm au}$ \\
${R_{\rm b}}^{\S}$ & $1490~{\rm au}$ & $1350~{\rm au}$ & $1910~{\rm au}$ & $1330~{\rm au}$ \\
${V_{\rm a}}^{\|}$ & $5300~{\rm km~s^{-1}}$ & $6480~{\rm km~s^{-1}}$ & $6010~{\rm km~s^{-1}}$ & $7610~{\rm km~s^{-1}}$ \\
${V_{\rm t}}^{\#}$ & $450~{\rm km~s^{-1}}$ & $600~{\rm km~s^{-1}}$ & $390~{\rm km~s^{-1}}$ & $710~{\rm km~s^{-1}}$ \\
$\displaystyle {\frac{dE_{\rm t}}{dt}}^{**}$&$2.2\times10^{35}~{\rm erg~s^{-1}}$
&$4.1\times10^{35}~{\rm erg~s^{-1}}$&$1.8\times10^{35}~{\rm erg~s^{-1}}$ & $5.8\times10^{35}~{\rm erg~s^{-1}}$\\
\hline
\end{tabular}\\
\end{center}
$^{*}$ Pericentric distance.\\
$^{\dagger}$ Eccentricity.\\
$^{\ddagger}$ Length of semi-major axis. \\
$^{\S}$ Length of semi-minor axis. \\
$^{\|}$ $V_{\rm a}$ is evaluated with the Sgr A* mass of
$4.31\times10^6~M_{\odot}$ \citep{Gillessen+2009}.\\
$^{\#}$ Eq. \ref{eq:tangential_velocity} is used.\\
$^{**}$ Eq. \ref{eq:tidalheating} is used. The duration time is fixed as one
year.\\
\end{table*}

\subsection{Summary} \label{sec:Summary}

We have performed fully three-dimensional simulations of the evolution of the G2
cloud. Our result differs from the result of previous two-dimensional
simulations in (i) strong vertical compression leads to the heating up and
flaring up of the cloud at the first pericenter passage, and (ii) because of
this compression, ram-pressure drag from the hot atmosphere is ineffective in
removing the energy and angular momentum of the cloud.

In our standard model, the peak luminosity would reach $\sim 100$ times the
solar luminosity. The luminosity depends on the assumed internal velocity
structure of the cloud, and thus might be fainter by one order of magnitude.
Since the peak luminosity is from the tidal compressional heating, it does not
depend on the assumption on the structure of the hot atmosphere around the SMBH.
We therefore believe that our prediction is pretty robust.

The increase of the luminosity of the cloud would be detectable in near infrared
bands about six months before the pericenter passage epoch.  In parallel, the
increase of the vertical velocity is probably observable as the line broadening.
Since the vertical velocity should strongly depend on the position of the gas on
the orbit, it is very important to measure the variation  of the velocity
profile both in space and time.

Detailed comparison between high-accuracy three-dimensional calculations and
observation will help us to understand the nature of the cloud and how it will
interact with the SMBH.

\bigskip
We thank the anonymous referee for his/her insightful comments and suggestions,
which helped us to greatly improve our manuscript.  Numerical simulations were
carried out on the Cray XT4 and XC30 systems in CfCA at NAOJ. This work is
supported by SPIRE.

\end{document}